\documentclass[10pt,twocolumn]{article} 
\usepackage{latex8}
\usepackage{times}

\usepackage[latin9,latin1]{inputenc}
\usepackage[T1]{fontenc}

\usepackage{graphicx,shadow,url}
\usepackage[english]{babel}
\usepackage{amsmath,amssymb}
\def\figdir{.}
\usepackage{listings,color}

\newcommand{\degre}{\ensuremath{^\circ}}

\begin{document}
\selectlanguage{english}

\title{Graphic processors to speed-up simulations \\ for the design of
  high performance solar receptors%
  \thanks{This work has been partially funded by the EVA-Flo project
    of the ANR and a STICS-UM2 multidisciplinary grant awarded to
    LIRMM, ELIAUS and PROMES laboratories.  This work has been
    possible thanks to the kind help of G.~Flamant, P.~Neveu,
    X.~Py and R.~Olives from PROMES laboratory (CNRS) and
    F.~André from CETHIL (CNRS-INSA Lyon).}%
}

\author{Sylvain Collange$^{\text{a}}$, Marc Daumas$^{\text{a,b}}$ and David Defour$^{\text{a}}$ \\
  $^{\text{a}}$ ELIAUS, UPVD --- $^{\text{b}}$  LIRMM, CNRS, UM2 \\
  52 avenue Paul Alduy --- Perpignan 66860 --- France \\
  firstname.lastname@univ-perp.fr \\
}

\maketitle
\thispagestyle{empty}

\begin{abstract}
  Graphics Processing Units (GPUs) are now powerful and flexible
  systems adapted and used for other purposes than graphics
  calculations (General Purpose computation on GPU --- GPGPU). We
  present here a prototype to be integrated into simulation codes that
  estimate temperature, velocity and pressure to design next
  generations of solar receptors. Such codes will delegate to our
  contribution on GPUs the computation of heat transfers due to
  radiations.  We use Monte-Carlo line-by-line ray-tracing through
  finite volumes.  This means data-parallel arithmetic transformations
  on large data structures.  Our prototype is inspired on the source
  code of GPUBench.  Our performances on two recent graphics cards
  (Nvidia 7800GTX and ATI RX1800XL) show some speed-up higher than 400
  compared to CPU implementations leaving most of CPU computing
  resources available. As there were some questions pending about the
  accuracy of the operators implemented in GPUs, we start this report
  with a survey and some contributed tests on the various floating
  point units available on GPUs.
\end{abstract}

\Section{Introduction}

Graphics Processing Units (GPU) offer computing resources higher than
the ones available on processors \cite{Pharr2005}. With the delivery
of the latest generations of GPUs, they can be used for general
processing (GPGPU, \url{www.gpgpu.org}) \cite{Man05} and become
application specific co-processors for regular and heavily
data-parallel processing.

We strongly believe that the development of GPGPU will necessary pass
through the identification of key applications that will benefit from
the various hardwired functionalities available on GPU.  We describe
the architecture of GPUs and properties of the implemented floating
point arithmetic discovered with our tests in Section~\ref{sec:pipeline_graphique}.
Section~\ref{sec/ray} presents the accurate estimation of radiative
heat transfers due to the filtering of incidental lines and the
generation of heat induced lines.  We elaborate on the performances of
our prototype and we present perspectives in Section~\ref{impl:res}.
We do not account for diffusion in this preliminary study as our
medium does not contain particles.

To the best of our knowledge, there is no prior art in the
implementation of the tasks reported here on GPUs. Monte-Carlo
ray-tracing and line-by-line analysis are routinely performed on CPUs
for simulations of radiative heat transfers though these tasks usually
saturate CPUs leaving no opportunity to the coupling of convective and
radiative phenomena on real applications. Other applications heavily
rely on elaborate physical models \cite{IbgHar02,JenRipWraJosElH07}.
Most simulations are currently performed for simple reference cases
(isothermal gas column at equilibrium). The description of gas
spectrum is generally simplified in calculation with engineering
interests leading to errors in the range of 5-15\%

\Section{Graphics Processing Units (GPU)}
\label{sec:pipeline_graphique}

\begin{figure}
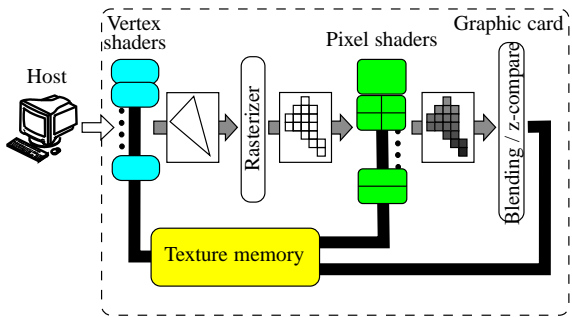

  \begin{center}
    \input graphics_pipeline_new_eng.pstex_t
  \end{center}
  \vspace{-12pt}
  \caption{Model of the graphics pipeline. }
  \label{fig:graphics_pipeline}
\end{figure}

GPUs handle mostly geometrical objects and pixels. Images are created
by applying geometrical transformations to vertices and by splitting
objects into pixels.  Calculations are carried out by
various stages composing the graphics pipeline presented in
Figure~\ref{fig:graphics_pipeline}. Actual pipelines of existing GPUs
differ slightly. Manufacturers move, share, duplicate or add resources
depending on boards and processors.  The figure shows the various
stages on the example of a triangle. In this example, vertex shaders treat 3 vertices
whereas pixel shaders treat 17 pixels. For most geometrical objects,
the number of pixels is  larger than the number of vertices.
Modern architectures contain more pixel shaders than vertex shaders.
The current ratio is commonly 24 against 8.
      
The host sends vertices to position primitive geometrical objects
(points, lines, polygons).  Objects are transformed (rotation,
translation, illumination\ldots) and assembled to create more
elaborate objects.  These operations are carried out by {\em vertex
  shaders}.

\begin{figure}
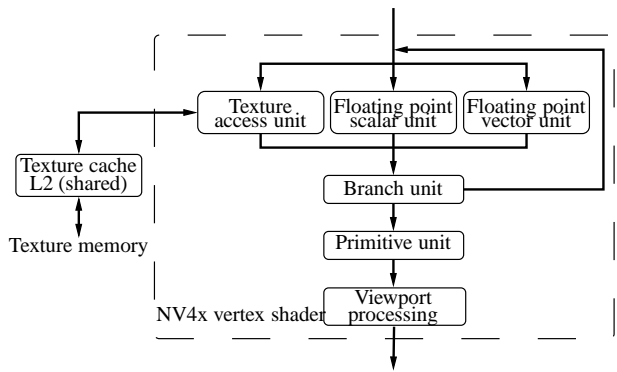

  \begin{center}
    \input vertexshader_new_eng.pstex_t
  \end{center}
  \vspace{-12pt}
  \caption{Vertex shader of the Nvidia 7800GTX.}
  \label{fig:vertexshader}
\end{figure}

At each cycle, each vertex shader (see Figure~\ref{fig:vertexshader}
adapted from \cite{Pharr2005}) is able to initiate a {\em Multiply and
  Add} (MAD) operation on 4 pieces of data in the vector unit
and a {\em special} operation in the scalar unit.  The implemented
{\em special} operations are exponential functions (exp, log),
trigonometric functions (sin, cos) and reciprocal functions ($1/x$ and
$1/\sqrt{x}$).  Since hardware support of DirectX 9.0, vertex shaders
are able to address texture memory through a dedicated unit.

\begin{figure}
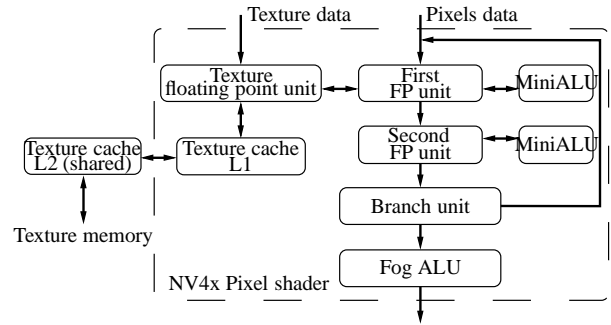

  \begin{center}
    \input pixelshader_new_eng.pstex_t
  \end{center}
  \vspace{-12pt}
  \caption{Pixel shader of the Nvidia 7800GTX. }
  \label{fig:pixelshader}
\end{figure}

The first floating point unit of each pixel shader (see
Figure~\ref{fig:pixelshader} adapted from \cite{Pharr2005}) carries
out 4 MADs or an access to texture via a dedicated unit.  The result
is then sent to the second floating unit which carries out 4 MADs.  In
the case of Nvidia 7800 GTX, each pixel shader includes a level 1
texture cache.

Table~\ref{fpformat} presents the floating point formats implemented
on GPUs and CPUs.  A number is represented by its mantissa, its exponent
$e$ and its sign bit $s$.  The first bit of the mantissa (left of the
fraction point) can be set to 1 unless the number to be represented is
very small.  The remaining bits form the fraction $f$. A normal
representation stores $(-1)^s ~ \cdot ~ 1 . f ~ \cdot ~ 2^e$ and a
subnormal one stores $(-1)^s ~ \cdot ~ 0 . f ~ \cdot ~ 2^{e_{\min}}$
where $e_{\min}$ is the minimum allowed exponent.  Single precision
(32 bit) became available on GPUs with Shader Model 3.0.
Manufacturers  of GPUs do not claim full compatibility with ANSI-IEEE
standard on floating point arithmetic.

Before porting our application to GPUs, we surveyed
two pieces of software testing performances and implementations of
floating point arithmetic on Nvidia 7800GTX and ATI RX1800XL
\cite{BucFatHan04, Hillesland2004}.  Tests have drawn the first
following conclusions:
\begin{itemize}
\vspace{-6pt}\item Additions and multiplications are truncated.
\vspace{-6pt}\item Subtractions seem to benefit from a guard bit with Nvidia but
  not with ATI.
\vspace{-6pt}\item Multiplications attain faithful rounding.
\vspace{-6pt}\item Errors on divisions indicate that divisions are based on
  multiplications by  approximations of the reciprocal.
\vspace{-6pt}\end{itemize}

\begin{table*}
  \caption{Representation format of floating point numbers on GPUs and CPUs.}
  \label{fpformat}
  \begin{center}
    \begin{tabular}{|l|c|c|c|c|c|}\hline
      Reference                   & \multicolumn{4}{|c|}{Number of bits}   & Non numerical  \\ \cline{2-5}
      & \parbox{15mm}{\centering Total} & \parbox{15mm}{\centering Sign} & \parbox{15mm}{\centering Exponent} & \parbox{15mm}{\centering Fraction} & values    \\ \hline \hline
      Nvidia                      & 16 & 1 & 5  & 10 & NaN, Inf                                \\ \cline{2-5}
      & 32 & 1 & 8  & 23 & (as documented in \cite{Cebenoyan2005}) \\ \hline
      ATI                         & 16 & 1 & 5  & 10 & Not implemented                         \\ \cline{2-5}
      & 24 & 1 & 7  & 16 &                                         \\ \cline{2-6}
      & 32 & 1 & 8  & 23 & Not documented                          \\ \hline \hline
      ANSI-IEEE 754 \cite{Ste.87} & 32 & 1 & 8  & 23 & NaN, Inf                                \\ \cline{2-5}
      & 64 & 1 & 11 & 52 &                                         \\ \hline
    \end{tabular}
  \end{center}
\end{table*}

\begin{table*}
  \caption{Arithmetic experimentations and results.}
    \label{tab/vect}
  \begin{center}
    \begin{tabular}{r|c|c|ll|} \cline{2-5}
         & Operations                                      & Shader  & \multicolumn{2}{c|}{Observations}                         \\ \cline{2-5} \cline{2-5}
         & $(M \oplus M) \ominus M$                        & All   & \multicolumn{2}{l|}{$M = 2^{127}(2-2^{-23}) \longrightarrow \infty$}            \\ \cline{2-5}
         & $\mathit{MAD}(x, y, - x \otimes y)$             & All   & \multicolumn{2}{l|}{$x \sim U [1, 2) \land y \sim U [1, 2) \longrightarrow 0$                       } \\ \cline{2-5}
         &                                                 &       & $1 \le i \le 23$    & $\longrightarrow 1.5 - 2^{-i}$     \\ %
         &                                                 & ATI-Pixel & $i = 24$            & $\longrightarrow 1.5 - 2^{-23}$    \\ %
         &                                                 &       & $25 \le i$          & $\longrightarrow 1.5$              \\ \cline{3-5}
         &                                                 &       & $1 \le i \le 23$    & $\longrightarrow 1.5 - 2^{-i}$     \\ %
         &                                                 & Nvidia-Pixel  & $24 \le i \le 25$   & $\longrightarrow 1.5 - 2^{-23}$    \\ %
         & $ 1.5 \ominus 2^{-i}             $              &       & $26 \le i$          & $\longrightarrow 1.5$              \\ \cline{3-5}
         &                                                 & ATI-Vertex & $1 \le i \le 23$    & $\longrightarrow 1.5 - 2^{-i}$     \\ %
         &                                                 &       & $24 \le i$          & $\longrightarrow 1.5-2^{-23}$      \\ \cline{3-5}
         &                                                 &       & $1 \le i \le 23$    & $\longrightarrow 1.5 - 2^{-i}$     \\ %
         &                                                 & Nvidia-Vertex  & $i = 24$            & $\longrightarrow 1.5 - 2^{-23}$    \\ %
         &                                                 &       & $25 \le i$          & $\longrightarrow 1.5$              \\ \cline{2-5}
         &                                                 &       & $1 \le i \le 23$    & $\longrightarrow 1.5 - 2^{-i}$     \\ %
         & $ (1 \oplus 0.5) \ominus 2^{-i}$                & All-Pixel & $24 \le i \le 25$   & $\longrightarrow 1.5 - 2^{-23}$    \\ %
         &                                                 &       & $26 \le i$          & $\longrightarrow 1.5$              \\ \cline{2-5}
         &                                                 &       & $1 \le i \le 23$    & $\longrightarrow      -2^{-i}$     \\ %
         &                                                 & ATI-Pixel & $i = 24$            & $\longrightarrow       -2^{-23}$   \\ %
         &                                                 &       & $25 \le i$          & $\longrightarrow 0$                \\ \cline{3-5}
         & $(1.5 \ominus 2^{-i}) \ominus 1.5$              &       & $1 \le i \le 23$    & $\longrightarrow      -2^{-i}$     \\ %
         &                                                 & Nvidia-Pixel  & $24 \le i \le 25$   & $\longrightarrow - 2^{-23}$        \\ %
         &                                                 &       & $26 \le i$          & $\longrightarrow 0$                \\ \cline{2-5}
         & $x \otimes y + (\pm x) \otimes (\mp y)$         & All   & \multicolumn{2}{l|}{$x \sim U [1, 2) \land y \sim U [1, 2) \longrightarrow 0$                       } \\ \cline{2-5}
         & $x \otimes y - (-x) \otimes (-y)$               & All   & \multicolumn{2}{l|}{$x \sim U [1, 2) \land y \sim U [1, 2) \longrightarrow 0$                       } \\ \cline{2-5}
         & $x \otimes y - ((2 \cdot x) \otimes y)/2$       & All   & \multicolumn{2}{l|}{$x \sim U [1, 2) \land y \sim U [1, 2) \longrightarrow 0$                       } \\ \cline{2-5}
         &                                                 & ATI-Pixel & \multicolumn{2}{l|}{$i \le (2^{11} - 1) \cdot 2^{12} \longrightarrow \textrm{correct}$}                       \\ \cline{3-5}
         & $(1 + 2^{-23}) \otimes (1 + 2^{-23} i)$         & Nvidia-Pixel  & $i \le 23 \cdot 2^{17}$ & $\longrightarrow \textrm{correct}$  \\ \cline{3-5}
         &                                                 & ATI-Vertex & $i \le 2^{23}$          & $\longrightarrow \textrm{correct}$  \\ \cline{3-5}
         &                                                 & Nvidia-Vertex  & $i \le 2^{19}$          & $\longrightarrow \textrm{correct}$  \\ \cline{2-5}
         &                                                 & ATI-Pixel & \multicolumn{2}{l|}{$x \in [1, 2) \land y \in [1,   2/x)  \longrightarrow \{-1.00031\textrm{ ulp} \cdots 0.00215\textrm{ ulp}\}$} \\ %
         &                                                 &       & \multicolumn{2}{l|}{$x \in [1, 2) \land y \in [2/x, 2  )  \longrightarrow \{-1.00013\textrm{ ulp} \cdots 0.00085\textrm{ ulp}\}$} \\ \cline{3-5}
         &                                                 & Nvidia-Pixel  & \multicolumn{2}{l|}{$x \in [1, 2) \land y \in [1,   2/x)  \longrightarrow \{-0.51099\textrm{ ulp} \cdots 0.64063\textrm{ ulp}\}$} \\ %
         & $x \otimes y - x \times y$                      &       & \multicolumn{2}{l|}{$x \in [1, 2) \land y \in [2/x, 2  )  \longrightarrow \{-0.76504\textrm{ ulp} \cdots 0.32031\textrm{ ulp}\}$} \\ \cline{3-5}
         &                                                 & ATI-Vertex & \multicolumn{2}{l|}{$x \in [1, 2) \land y \in [1,   2/x)  \longrightarrow \{-1\textrm{ ulp} \cdots 0\}$} \\ %
         &                                                 &       & \multicolumn{2}{l|}{$x \in [1, 2) \land y \in [2/x, 2  )  \longrightarrow \{-1\textrm{ ulp} \cdots 0\}$} \\ \cline{3-5}
         &                                                 & Nvidia-Vertex  & \multicolumn{2}{l|}{$x \in [1, 2) \land y \in [1,   2/x)  \longrightarrow \{-0.82449\textrm{ ulp} \cdots 0.93750\textrm{ ulp}\}$} \\ %
         &                                                 &       & \multicolumn{2}{l|}{$x \in [1, 2) \land y \in [2/x, 2  )  \longrightarrow \{-0.91484\textrm{ ulp} \cdots 0.46875\textrm{ ulp}\}$} \\ \cline{2-5}
    \end{tabular}
  \end{center}
\end{table*}

We wrote additional test vectors summarized in Table~\ref{tab/vect}
where $\oplus$, $\ominus$, $\otimes$ are the addition, subtraction and
multiplication operators implemented on GPU.  $U[a, b)$ are uniformly
distributed random variables on $[a, b)$. Random tests are performed
on $2^{23}$ inputs, other tests are exhaustive.

We used OpenGL primitives and stored data in textures using {\em Frame
  Buffer Object} and respectively {\em texRECT} and {\em tex2D}
   for Nvidia and ATI chips. We set up vertex and pixel shaders for
computation with the OpenGL shading language. We ran these tests on
Nvidia 7800 GTX with driver ForceWare 81.98 and on ATI RX1800XL with
driver Catalyst 6.3.

\begin{figure*}
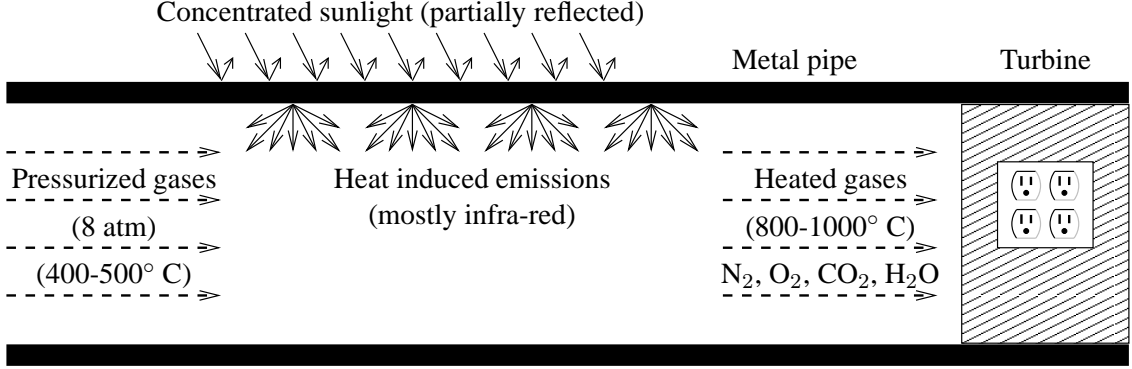

  \begin{center}
    \input setting.pstex_t
  \end{center}
  \vspace{-12pt}
  \caption{The solar receptor as simulated.}
  \label{fig:setting}
\end{figure*}

On some architectures, internal registers store numbers with a
precision higher than the one used in memory or with a larger dynamics
for the exponents. Sometimes MADs maintain larger accumulators or
round results only once, after the addition.  Our tests showed that no
such things occur on GPUs but they revealed a surprising feature.
It appears that the second pixel shader floating point unit on ATI and
both units on Nvidia produce a mantissa with
one extra bit.
This extra bit forces modifications of some multiple
precision tools \cite{DaGDef06} and we conjecture that it is
implemented for backward compatibility.

Fast small multipliers usually ignore partial products below a given
threshold and add a constant to correct the introduced statistical
bias \cite{SchSwa93b}. Results lead us to think that this constant is
$2^{-35}$ on ATI and $41 \cdot 2^{-30}$ on Nvidia. The multipliers
accumulate partial products on 9 extra rows on ATI and 6 extra rows on
Nvidia. These figures do not include the extra bit left of the
mantissa.  Other tests indicate that multipliers use radix 2
sign-magnitude logic internally.

Additional tests showed that subnormal numbers are replaced by 0
during transfers even when no arithmetic operation is performed on GPUs
meaning that drivers probably perform arithmetic operations on
textures. Non numerical quantities are not modified except that sNaN
({\em signaling NaN}) is changed to qNaN ({\em quiet NaN}) on ATI.

\Section{Monte-Carlo line-by-line ray tracing}
\label{sec/ray}

The experimental setting is presented in Figure~\ref{fig:setting}.
This device produces electricity from sunlight concentrated by a large
reflector.  Concentrated sunlight is used to heat a metal pipe that
transfers heat through contact and infra-red radiations.  The goal is
to transfer as much energy as possible to the turbine.  Dynamic and
thermal phenomena are intricately interwoven as air temperature
raises.

Though our approach is based on finite volumes used for example by
Fluent (\url{www.fluent.com}) and Trio-U (\url{www-trio-u.cea.fr}),
this work can also be applied to accurately instantiate source terms
in software based on finite element methods such as ComSol
(\url{www.comsol.fr}).

Combined optical depth $\tau$ of infrared participating gases CO$_2$ and H$_2$O
(O$_2$ and N$_2$ are ignored) represents millions of lines that are
functions of temperature $T$, pressure $p$, and density of absorbing
molecule $u_g$ in the following formulas copied from \cite[Annex
A.2]{Rot.98} with the same notations.  
\begin{eqnarray}
  \label{eqn/S} \frac{S_{\eta \eta'}(T)}{S_{\eta \eta'}(T_{\text{ref}})} & =  &
    \frac{Q(T_{\text{ref}})}{Q(T)}
    \frac{e^{-\frac{c_2E_{\eta}}{T}}}{e^{-\frac{c_2E_{\eta}}{T_{\text{ref}}}}}
    \frac{\left(1 - e^{-\frac{c_2\nu_{\eta \eta'}}{T}}\right)}{\left(1 - e^{-\frac{c_2\nu_{\eta \eta'}}{T_{\text{ref}}}}\right)} \\
  \label{eqn/tau} \tau(\nu)      & = &  \sum_{g} u_g \sum_{\eta \rightarrow \eta'} S_{\eta \eta'}(T)f (\nu - \nu_{\eta \eta'})   \\
  \label{eqn/I} I_{\text{out}} & = & I_{\text{in}} e^{- \tau (\nu) l} + I(\nu, T) \left(1 - e^{- \tau (\nu) l}\right)            \\
  \label{eqn/planck}I(\nu, T) & = & \frac{2h\nu^3}{c^2} \cdot \frac{1}{\left(e^{\frac{c_2 \nu}{T}} - 1\right)}
\end{eqnarray}

\begin{figure}
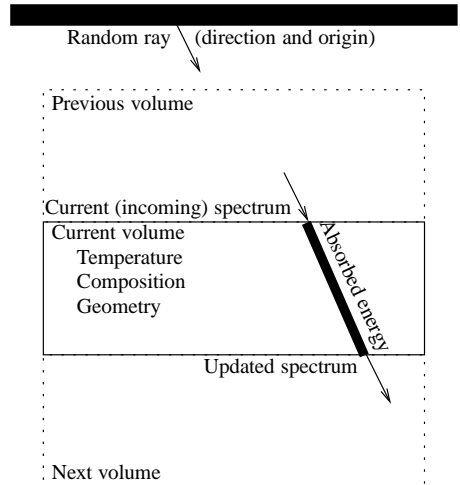

  \begin{center}
    \input volume.pstex_t
  \end{center}
  \vspace{-12pt}
  \caption{Monte-Carlo ray-tracing.}
  \label{fig:ray}
\end{figure}

The first formula provides a ratio $S_{\eta \eta'}(T)/S_{\eta
  \eta'}(T_{\text{ref}})$ for the intensity of the line due to
transition between lower and upper states $\eta$ and $\eta'$ of
component gas $g$ centered on wavenumber $\nu_{\eta \eta'}$. This
ratio is applied to the 16 contributions of this line in the
wavelength space around $\nu_{\eta \eta'}$. Once this transformation
is performed for all the lines of all the gases, the contributions are
cumulated to obtain $\tau({\nu})$ for all the considered wavenumbers
$\nu$. We apply Beer-Lamber law for absorption (first term of
$I_{\text{out}}$) and Planck law for heat induced emissions (second
term of $I_{\text{out}}$) for a ray passing through length $l$ of an
isothermal homogeneous finite volume of Figure~\ref{fig:ray}.  GPUs
handle all data-parallel computations and Listing~\ref{lst/trans}
presents how formulas (\ref{eqn/tau}), (\ref{eqn/I}) and
(\ref{eqn/planck}) are implemented.

\begin{lstlisting}[float,{basicstyle=\ttfamily\footnotesize},caption=Parallel evaluation of (\ref{eqn/tau})-(\ref{eqn/planck}),label=lst/trans,language={sh}]
!!ARBfp1.0

...
# (*@$\texttt{sratio\_g\{1-2\}} = S_{\eta \eta'}(T)/S_{\eta \eta'}(T_{\text{ref}})$@*) are computed
# in the omitted part from ((*@{\tt\it \ref{eqn/S}}@*))
# and stored in a texture
TEX sratio_g1, sig_coords_1, texture[4], RECT;
TEX sratio_g2, sig_coords_2, texture[4], RECT;

# (*@$\texttt{tref\_g\{1-2\}} = u_g S_{\eta \eta'}(T_{\text{ref}}) f (\nu - \nu_{\eta \eta'})$@*) are constant
# textures computed on CPU and transfered to 
# GPU memory on program initialization
TEX tref_g1, vnu_coords_1, texture[5], RECT;
TEX tref_g2, vnu_coords_2, texture[5], RECT;

# Suming up the contributions of the two gases
MUL tau, sratio_g1, tref_g1;
MAD tau, sratio_g2, tref_g2, tau;

MUL tau, tau, ll;
# (*@$\texttt{ll} = - l / \ln (2)$@*) is a scalar
# set for each iteration
# Factors $1/ln(2)$ are introduced as GPUs
# currently only support base-2 exponentials

# Special functions need 4 invocations
EX2 exp_tau_l.x, tau.x;
EX2 exp_tau_l.y, tau.y;
EX2 exp_tau_l.z, tau.z;
EX2 exp_tau_l.w, tau.w;

MUL exponent, c2T, nu;
# (*@$\texttt{c2T} = c_2 / (T \ln (2))$@*)  is a scalar
# set for each iteration

EX2 den.x, exponent.x;
EX2 den.y, exponent.y;
EX2 den.z, exponent.z;
EX2 den.w, exponent.w;

SUB den, den, {1, 1, 1, 1};

RCP inv.x, den.x;
RCP inv.y, den.y;
RCP inv.z, den.z;
RCP inv.w, den.w;

MUL nu3, nu, nu;
MUL nu3, nu3, nu;
MUL nu3, nu3, hc2;
# (*@$\texttt{hc2} = 2h /c^2$@*) is a constant scalar


MUL factor1, inv, nu3;
SUB factor2, one, exp_tau_l;
MUL term, i_in, exp_tau_l;
# (*@$\texttt{i\_in} = I_{\text{in}}$@*) is the (*@$I_{\text{out}}$@*) texture of
# the previous iteration

# Return (*@$\texttt{result.color} = I_{\text{out}}$@*) as the color of
# the pixel to be written
MAD result.color, factor1, factor2, term;

END
\end{lstlisting}

After the GPU has computed $I_{\text{out}}$ for all the considered
wavenumbers the power of the total heat transfer is obtained by
summing $I_{\text{in}} - I_{\text{out}}$ of up to $2^{24} \approx 16
\cdot 10^6$ values stored in a 2 dimensional square matrix.  This task
requires to sum all the data of a texture and we used a parallel
reduction scheme adapted to GPUs \cite{Pharr2005}. The sum is
evaluated with an iterative algorithm where each iteration sums of $4$
pieces of data from the previous iteration.

Integrations with respect to space in the simulation of non-isothermal
flows is obtained by Monte-Carlo line-by-line ray-tracing
paradigm as presented Figure~\ref{fig:ray}. The main simulation code
on CPU directs this process and averages the effect of individual
rays.

We designed a program in two parts. The first part is executed by the
CPU and represents 3500 lines of C++ code and OpenGL directives. The
second part is executed by the pixel shaders of the GPU and represents
250 lines of OpenGL shading primitives (ARB fragment program).
Listing~\ref{lst/trans} is extracted from these 250 lines and
corresponds to the evaluation of formulas (\ref{eqn/tau}),
(\ref{eqn/I}) and (\ref{eqn/planck}).  In addition, we wrote the same
program in C to measure the benefit of the use of a GPU. This code was
compiled with Microsoft Visual C++ 2005, optimizing for speed
(\verb+/Ox /arch:SSE+).  We ran both programs on a Pentium 4 system
with 1~GB of DDR2 memory and with a Nvidia 7800GTX and an ATI RX1800XL
graphic card both with 256~MB of GDDR3.

\begin{figure}
  \begin{center}
    \includegraphics[width=1.1\columnwidth]{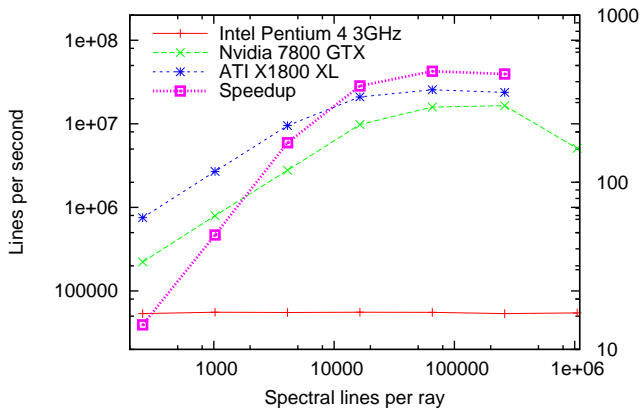}
  \end{center}
  \vspace{-12pt}
  \caption{Number of lines treated by second.}
  \label{fig:speed}
\end{figure}

We measured the number of lines evaluated per second depending on the
number of lines per ray. The results are plotted in
Figure~\ref{fig:speed} with logarithmic axes and show a speed-up as
high as 420 compared to CPU.  Timing is done on 100 rays treated
sequentially.  GPU performance loss around $10^6$ lines per ray is due
to data too large to fit in graphic memory and should disappear with
newer GPU boards.

This impressive speed-up is partially due to the ability of GPUs to
perform many complex operations per cycle.  Each pixel shader can
start one exponential per cycle thanks to dedicated hardware. As there
are up to 24 shaders, 24 exponentials are initiated at 486 Mhz leading
to $13.2~10^9$ exponentials per second. On CPUs, exponential
functions are evaluated in software or in micro-code and require
typically $100$ cycles to complete. On a 3~Ghz Pentium 4 this means
about $30~10^6$ exponentials per second. The second reason for our
speed-up lies in the fact that GPUs and drivers exploit regularity in the
code to hide memory latency and execute floating point operations in
parallel in pixel shaders.
 
\Section{Conclusion and perspectives}
\label{impl:res}

We started this report with test vectors aimed at the characterization
of the floating point operators of GPUs. We showed that:
\begin{itemize}
\vspace{-6pt}\item Temporary results are computed to 32 bit format.
\vspace{-6pt}\item Multipliers use constants to compensate for discarded partial
  products.
\vspace{-6pt}\item Some Nvidia and ATI adders use an extra bit.
\end{itemize}
We will certainly set up more test vectors as we continue working on
GPUs. Up to date tests are available from the authors upon request by
email.

We accelerated the computation of radiation properties in order to
simulate precisely, i.e. using line-by-line spectra of gases.
Common speed-up brought by GPU start at 5 and may climb to 50 as some 
developments in the industry are claiming\footnote{See
  \url{http://www.emphotonics.com/fastfdtd.html}.}. 
Our GPU implementation is 400 times faster than CPU evaluation.
This performance almost preserves the computing resource available on
CPU as we noticed a runtime increase below 1\%
program saturates our CPU and GPU compared to the same program with no
request to GPU.

These figures where obtained using a fixed number (16) of points of
evaluation for each line. Our next version will dynamically adapt the
number of points depending on the local temperature and the intensity
of the line. This tasks will involve vertex shaders and blending
units. Blending units starting with Nvidia 8800 operate on 32 bit
floating point data.  Work on radiosity will be performed only if
discrepancies between simulations and experimentations show that the
effect of diffusion cannot be ignored.

The impressive speed-up reported here was due to the large number of
lines for one single ray-tracing leading to a huge amount of data
parallel transformations. Similar speed-ups may be obtained for other
settings. One possible application of GPGPU with connection to the
industry, is to speed-up simulations of elaborate surfaces of planar
solar receptors.  Software will average spectral effects to two bands
of wavelength (infrared and visible) but it will consider a
large number of independent rays to accurately account for anisotropic
reflections and absorptions.

As we are building know-how on porting simulations for thermal
sciences to GPUs we will explore automatic tools and build libraries
of techniques to efficiently reuse parts of our developments.

\bibliographystyle{latex8}
\bibliography{alternate,daumas_ref}

\end{document}